\documentclass[12pt]{iopart}

\usepackage{iopams}
\usepackage{hyperref}
\hypersetup{colorlinks=true, urlcolor=red, citecolor=blue, linkcolor=blue}
\usepackage{graphicx}
\usepackage{subfigure}
\usepackage{epstopdf}
\usepackage{epsfig}
\usepackage{newlfont}
\usepackage{amssymb}
\usepackage{amsfonts}

\begin{document}

\renewcommand*{\DefineNamedColor}[4]{%
   \textcolor[named]{#2}{\rule{7mm}{7mm}}\quad
  \texttt{#2}\strut\\}

\definecolor{red}{rgb}{1,0,0}
\definecolor{cyan}{cmyk}{1,0,0,0}

\title[Mapping generalized Jaynes-Cummings interaction...]{Mapping generalized Jaynes-Cummings interaction into correlated 
finite-sized systems}

\author{Himadri Shekhar Dhar\(^1\), Arpita Chatterjee\(^1\) and Rupamanjari Ghosh\(^{1,2}\)}

\address{\(^1\)School of Physical Sciences, Jawaharlal Nehru University, New Delhi 110067, India\\
\(^2\)School of Natural Sciences, Shiv Nadar University, Gautam Buddha Nagar, UP 201314, India
}
\ead{rghosh.jnu@gmail.com}

\begin{abstract}
We consider a generalized Jaynes-Cummings model of a two-level atom interacting with a multimode nondegenerate 
coherent field. The sum of the mode frequencies is equal to the two-level transition frequency, creating the 
resonance condition. The intermediate levels associated with the multi-photon process are adiabatically eliminated using 
the non-resonant conditions for these transitions. Under such general conditions, the infinite atom-multiphoton 
interaction is effectively mapped onto an equivalent reduced \textit{2}$\times$\textit{2} bipartite qubit system that 
facilitates the study of the nonclassical features of the interaction using known information-theoretic measures. We 
observe that the bipartite pure system is highly entangled as quantified by its entanglement of formation. Further, it is 
shown that the dynamics of the mapped system can be generated using optically truncated, quantum scissored states that reduce the infinite atom-multiphoton interaction to a finite \textit{2}$\times$\textit{k} system, where $k$ is a suitable truncation number. This allows us to introduce atomic dephasing and study the mixed state dynamics, characterized by the decay of quantum correlations such as quantum discord, 
which is observed to be more robust than 
entanglement. 
The quantum correlation dynamics of the dissipative system qualitatively complements the behavior of collapse and revival of the Rabi oscillations in the system. The effective mapping of the composite system proves to be an efficient tool for measuring information-theoretic properties.
\end{abstract}

(Some figures may appear in color only in the online journal)



\submitto{\JPB}
\maketitle

\section{Introduction}
\label{intro}

The archetypal work horse representing the simplest quantum interaction, that of a single atom and a single cavity mode, 
is the Jaynes-Cummings (JC) model \cite{JC}. The model was first proposed to study spontaneous emission and examine the 
underlying semiclassical behavior of quantum radiation \cite{Shore&Knight}. The periodic collapse and subsequent revival 
of Rabi oscillations \cite{Eberly} in the temporal dynamics of the JC model later proved the discrete and quantum nature 
of photons. In the following years, the JC interaction has been generalized to dissipative and perturbative models 
\cite{gen} to study various fundamental aspects of the manifested quantum atom-photon interaction \cite{cohen}. 
Experimentally, the advances in cavity quantum electrodynamics involving Rydberg atoms in single-mode cavities 
\cite{Rempe, Haroche} have allowed investigation of quantum features and possible extension of the properties of 
atom-field interaction for the implementation of quantum tasks \cite{condense}.

The JC model is an extremely purposeful model from the perspective of quantum information theory (QIT) and quantum 
computation. The nonclassical interaction between the atomic levels and the cavity modes gives rise to  
entanglement \cite{Qent}, which is the primary resource for performing various QIT tasks such as superdense coding 
\cite{DC}, quantum teleportation \cite{tele}, and quantum key distribution \cite{key}. Coupled arrays of JC 
systems have been considered to generate and study correlated many-body entangled states \cite{Plenio}, which can be 
applied in quantum computing and networks \cite{net}. Further, the manipulation and control of entanglement in different 
generalized versions of JC model have received a lot of attention (for a review, see \cite{HarocheRMP}). However, the 
quest for suitable quantum systems to operationalize various aspects of QIT for practical purposes is still far from over. 
The main obstacles are the difficulty in the manipulation of quantum states and the irreversible decoherence present in 
most interacting quantum systems \cite{death}. With the advent of innovative experiments with ultracold atoms 
\cite{ultracold} and optical lattices \cite{Blatt}, it has been possible to increase interaction and manipulability of 
many-body quantum systems.

The most pervasive feature of the JC interaction is the nonclassical nature of the field and its correlation with the 
atomic levels. The dynamics of these atom-cavity correlations is the key to the interesting properties of all generalized 
JC models. For QIT applications, the quantum correlations need to be studied from the relevant perspective. From 
information-theoretic arguments, the concept of nonclassical correlations is neither limited to nor described fully by 
entanglement. The concept of measurement-based, information-theoretic quantum correlation measures, such as quantum discord (QD) \cite{discord} and quantum information deficit (QID) \cite{qid} use the fact that physical measurements disturb the noncommutative nature of quantum systems and thus erase quantum correlation \cite{luo}, and are, in general, non-monotonic with measures of 
entanglement such as entanglement of formation \cite{eof} and logarithmic negativity \cite{LN}. 
These information-theoretic measures are known to be non-zero for separable mixed states.
There exist non-trivial operational relations between entanglement and other quantum correlation measures such as QD and QID \cite{ref,ref1}, however, in general there is no established hierarchy between these measures \cite{ref2}. 
The information-theoretic measures assume importance when one considers the fact that efficient quantum tasks have been performed with states that have no entanglement \cite{zero} (for a review, see \cite{modi}). 
Further, it has been shown that QD is an optimal resource \cite{optimal} -- it is more robust \cite{robust} than entanglement and does not exhibit sudden death \cite{sudden}, though contrary results have also been reported \cite{steve}. Hence, these quantum correlations should be critically explored in systems that are designed to perform computation and other QIT tasks.

In this article, we seek to understand and characterize the behavior of 
two conceptually different quantum correlation measures, viz. entanglement and quantum discord, arising in the dynamics of a two-level atom interacting with a multimode coherent field in a generalized JC model \cite{JC}. 
In this interaction of a two-dimensional atomic system with an infinite-dimensional multimode radiation field, we take the 
sum of the mode frequencies to be equal to the effective two-level transition frequency, creating the resonance condition. 
The intermediate levels associated with the multi-photon process can then be adiabatically eliminated using the 
non-resonant condition for these transitions. The nature of the interaction allows us to map the pure composite system onto an 
equivalent reduced \textit{2}$\times$\textit{2} system. Such a mapping is known for single-mode input fields 
\cite{knight}. Thus the infinite-dimensional problem gets mapped into a bipartite qubit system that simplifies the 
correlation dynamics and allows the investigation of important characteristics of the quantum system in a reduced Hilbert 
space. 
Further, for dissipative, nonunitary evolution of the model under the influence of environment, the dynamics of the system can be studied using optically truncated, quantum scissored states \cite{sciss}. For unitary dynamics, we show that an optically truncated, reduced \textit{2}$\times$\textit{k} quantum system can simulate the pure state dynamics of the mapped two-qubit system. 
Under dissipative interactions in the atom-photon interaction, the optically truncated, finite-size system can be used to study the behavior of quantum correlation emerging from the mixed state dynamics of the composite open system.

Our general mapping scheme leading to the reduced dynamical structure of the problem can be used for studying important 
correlation properties or photon statistics in similar model systems with multimode fields. Using single- and two-mode 
fields, we observe that the reduced bipartite system is highly entangled, as quantified by its entanglement of formation. Hence, the interaction can be used to generate maximally entangled bipartite systems for information processing and communication. Under dissipative atomic dephasing, we study the temporal decay of the correlations, and QD turns out to be more robust than entanglement. 
Additionally, we observe that the quantum correlation dynamics complements the collapse and revival of the Rabi oscillations and its subsequent decay in the damped atom-field interaction. For the features in our investigation, the use of a multimode field gives qualitatively similar results.

This paper is organized as follows. We begin by defining the various quantum correlation measures in section 
\ref{correlation}. We describe our generalized JC model of an effective two-level atom interacting with a nondegenerate 
\textit{N}-mode coherent field in section \ref{system}. In section \ref{mapping}, we present 
the mapping of our composite system onto an equivalent \textit{2}$\times$\textit{2} system, and provide an alternate description using optically truncated, quantum scissored \textit{2}$\times$\textit{k} state. Section \ref{results} gives the dynamics of entanglement for the pure composite atom-field system with single- and two-mode fields. In section \ref{damp}, we discuss the effect of nonunitary atomic dephasing on quantum correlations. Section \ref{dissc} contains discussions of our main results.

\section{Quantum correlation measures}
\label{correlation}

For pure states, most entropic quantum correlation measures such as entanglement of formation and quantum discord
reduce to the entropy of entanglement \cite{eoe} whereas logarithmic negativity gives an upper bound. For mixed states, the operational relation between these measures is non-trivial \cite{ref, ref1}, and there is no unique characterization or hierarchy \cite{ref2}. In this section, we briefly define two entanglement measures, viz. entanglement of formation and logarithmic negativity, and the information-theoretic quantum correlation measure of quantum discord.  \\

\noindent\emph{Entanglement}:
The entanglement of formation (EOF) of a bipartite quantum state \cite{eof} is defined as the average entropy of 
entanglement of its pure state decomposition, optimized over all such possible decompositions. For a bipartite quantum 
state, $\rho = \sum_i p_i |\psi_i\rangle\langle \psi_i|$, the entanglement of formation is defined as
\begin{equation}
E(\rho)= \textrm{min}_{\{\phi\}} \sum_i p_i E'(|\psi_i\rangle) ,
\end{equation}
where $E'(|\psi\rangle)$ is the entropy of entanglement, defined as the von Neumann entropy of the subsystem of any 
bipartite pure state, $|\psi\rangle$. The minimization is over all possible pure state decompositions ($p_i, 
|\psi_i\rangle$) of the bipartite state, $\rho$. For pure states, the EOF reduces to the entropy of entanglement.
The optimization or convex-roof construction to obtain the mixed state EOF is not easily tractable. For a two-qubit system, the EOF has a closed analytical form defined in terms of its concurrence \cite{con}. For arbitrary dimensions, the convex-roof construction for mixed state EOF is not very well defined, though non-trivial generalizations of concurrence have been explored \cite{mintert}.  

A computable measure of entanglement for mixed systems is the logarithmic negativity (LN) \cite{LN}. The definition of LN stems from the fact that the negativity of the partial transpose of a bipartite quantum system is a sufficient condition for entanglement \cite{suff}. For two-qubit and ${\textit 2}\times{\textit 3}$ systems, this is also a necessary condition \cite{necc}. The LN of an arbitrary bipartite state, $\rho$, is defined as
\begin{equation}
L(\rho) = \log_2[2 N(\rho)+1] ,
\end{equation}
where $N(\rho)$ is called the negativity. It is defined as $N(\rho)= \frac{1}{2}(\left\|\rho^{T_A}\right\| - 1)$, where 
$\left\|\rho^{T_A}\right\|$ is the trace norm of the partially transposed state, $\rho^{T_A}$, of $\rho$. For pure states, the LN does not reduce to the entropy of entanglement and can be zero for bound entangled states in bipartite systems with dimensions higher than ${\textit 2}\times{\textit 3}$. However, non-zero values of LN is a known upper bound on distillable entanglement (for mixed states) or entropy of entanglement (for pure states). Due to its computational simplicity, LN is often used as a measure of distillable entanglement in higher dimensions. The significance of LN is also highlighted in establishing possible relation of distillable entanglement with information-theoretic measures such as quantum discord \cite{ref}. \\

\noindent\emph{Quantum discord}:
The definition of quantum discord \cite{discord} is borne from the intricate nature of measurement in quantum 
mechanics and its role in disturbing the noncommutative nature of quantum operators \cite{luo}. In classical information 
theory, the total correlation between two random variables is contained in the mutual information between these variables. 
The mutual information can be arrived at using the classical Shannon entropy via two distinct methods: first, using the 
idea of joint probability distribution, and secondly, using the idea of conditional entropy. Extending the concept to 
the quantum regime and replacing Shannon entropy with the von Neumann entropy, we obtain two distinct, inequivalent but 
classically equal definitions of mutual entropy:
\begin{eqnarray}
\label{mutual}
I(\rho(t))&\equiv& S(\rho_a)+ S(\rho_f)- S(\rho(t)), \mbox{and} \\
\tilde{I}(\rho(t)) & \equiv& S(\rho_f) - S(\rho_{f|a}),
\end{eqnarray}
where $S(\rho)= - \mbox{tr} (\rho \log_2 \rho)$ is the von Neumann entropy of the quantum state $\rho$. $\rho_a(t)$ and 
$\rho_f(t)$ are the reduced density operators of the atomic and the field subsystems, respectively. $I(\rho(t))$ is the quantum mutual information \cite{mut} and $S(\rho_{f|a})$ is the quantum conditional entropy \cite{cond}. Quantum discord is defined as the difference, $Q(\rho(t))=I(\rho(t))- \tilde{I}(\rho(t))$. In the classical regime, $I(\rho(t))\equiv 
\tilde{I}(\rho(t))$.

The inequivalence of the two expressions in the quantum regime is due to the quantum conditional entropy, $S(\rho_{f|a})$, 
which is defined as the lack of information about one subsystem (say, $\rho_f$) when that of the other ($\rho_a$) is 
known. The complete knowledge of a subsystem invokes a measurement on the subsystem.
Let $\{\mathcal{F}^i_a\}$ $(i = 1,2)$ be a basis set of one-dimensional projectors (with $\mathcal{F}^i_a \mathcal{F}^j_a$ = $\delta_{ij} \mathcal{F}^i_a$, $\sum_i \mathcal{F}^i_a$ = $\mathbb{I}_a$) acting on the two-dimensional atomic subsystem ($\rho_a$). 
$\mathbb{I}_{a}(\mathbb{I}_{f})$ is the identity operator acting on the Hilbert space on which that atomic (field) 
subsystem, $\rho_a$($\rho_f$), is defined. If one makes a projective measurement on the atomic subsystem, the 
post-measurement density matrix can be written as $\rho^i_{af}(t)= \frac{1}{p_i(t)}(\mathcal{F}^i_a \otimes\,\,\mathbb{I}_f ~ \rho(t)\, \mathcal{F}^i_a \otimes\, \mathbb{I}_f)$, where $p_i(t) = \mbox{tr}_{af}(\mathcal{F}^i_a \otimes\,\,\mathbb{I}_f ~ \rho(t)\, \mathcal{F}^i_a \otimes\, \mathbb{I}_f)$.
The quantum conditional entropy is then given by
\begin{equation}
\label{cond}
S(\rho_{f|a}) = \min_{\{\mathcal{F}^i_a\}} \sum_i p_i(t) S(\rho^i_{af}(t)),
\end{equation}
where the minimization is over all possible sets of rank-1 projective measurements, \{$\mathcal{F}^i_a$\}. 
$\tilde{I}(\rho(t))$ and QD can then be defined as
\begin{eqnarray}
\tilde{I}(\rho(t)) & = & S(\rho_f) - \min_{\{\mathcal{F}^i_a\}} \sum_i p_i(t) S(\rho^i_{af}(t)),\\
Q(\rho(t))& = & I(\rho(t)) -\tilde{I}(\rho(t)).
\label{QD}
\end{eqnarray}
The measurement involved in $\tilde{I}(\rho(t))$ ensures that $I(\rho(t)) \ge \tilde{I}(\rho(t))$. Hence, $Q(\rho(t)) \ge 
$0 and is also known to be non-zero for separable states. As mentioned, in our calculations we always apply the projective measurement on the two-dimensional atomic subsystem.\\

\section{Generalized interaction of a two-level atom with a multimode field}
\label{system}

We consider the interaction between an effective two-level atom and a nondegenerate \textit{N}-mode coherent field. The 
excited state and the ground state of the two-level atom are denoted by $|e\rangle$ and $|g\rangle$, respectively. The 
transition frequency between the levels $|e\rangle$ and $|g\rangle$ is denoted by $\omega_0$. $\omega_1$, $\omega_2$, ..., 
and $\omega_N$ are the frequencies of the \textit{N} modes of the field, where $\omega_1 + \omega_2 +...+\omega_N = 
\omega_0$. The intermediate states associated with any multi-photon process can be adiabatically removed for the 
considered frequencies with the additional assumption that the intermediate transition frequencies are not resonant with 
the chosen field modes \cite{zub}. In the rotating-wave approximation, the effective Hamiltonian 
is described by \cite{scully97} ($\hbar=1$)
\begin{eqnarray}
H &=& H_0 + H_{int} \nonumber \\
&=& \omega_0\sigma_z + \Sigma_{j=1}^N \omega_j a_j^\dag a_j+ \mathrm{\tilde{g}}(\cal{A}_\alpha\sigma^++ 
\cal{A}_\alpha^\dag \sigma^-) ,
\end{eqnarray}
where $\cal{A}_\alpha = \prod_{j=1}^N a_j$, $\cal{A}_\alpha^\dag = \prod_{j=1}^N a_j^\dag$. Further, $\sigma_z = 
\left|e\rangle\langle e\right| - \left|g\rangle\langle g\right|, \,\, \sigma^+=\left|e\rangle\langle g\right|, 
\,\,\sigma^-=\left|g\rangle\langle e\right|$, $a_j^\dag$ and $a_j$ are the creation and annihilation operators, 
respectively, of the field modes $j=1,2,...,N$, and $\mathrm{\tilde{g}}$ is the electric dipole coupling constant.
The unitary time-evolution operator derived from the above Hamiltonian \cite{scully97} is
\begin{eqnarray}\nonumber
U(t)&=&\exp(-i H_{int}t)\nonumber\\
&=& \cos(\hat{\cal{N}}_1 \mathrm{\tilde{g}}t)|e\rangle\langle e|-i\cal{A}_\alpha \frac{\sin(\hat{\cal{N}}_2 
\mathrm{\tilde{g}}t)}{\hat{\cal{N}}_2} |e\rangle\langle g| \nonumber\\
&& +\cos(\hat{\cal{N}}_2 \mathrm{\tilde{g}}t)|g\rangle\langle g|-i\cal{A}_\alpha^\dag\frac{\sin(\hat{\cal{N}}_1 
\mathrm{\tilde{g}}t)}{\hat{\cal{N}}_1}|g\rangle\langle e|,
\end{eqnarray}
where $\hat{\cal{N}}_1=\sqrt{\prod_{j=1}^N a_j a_j^\dag}$, and $\hat{\cal{N}}_2=\sqrt{\prod_{j=1}^N a_j^\dag a_j}$.

The initial ($t=0$) nondegenerate multimode infinite-dimensional coherent field can be written as \cite{roy}
\begin{eqnarray}
|\psi_f(0)\rangle &=&\prod_{j=1}^N|\alpha_j\rangle = \sum_{n_1,n_2,...,n_N=0}^\infty 
\cal{C}_{n_j}^{\alpha}\prod_{j=1}^N|n_j\rangle , \nonumber \\ \nonumber \\
 \cal{C}_{n_j}^{\alpha} &=& \exp\left(\frac{-\sum_{j=1}^N |\alpha_j|^2}{2}\right) 
 \prod_{j=1}^N\frac{\alpha_j^{n_j}}{\sqrt{n_j!}} ,
\end{eqnarray}
where $|\alpha_j\rangle$ is a single-mode coherent state with complex amplitude $\alpha_j$ \cite{roy,sud}.

The initial state of the two-level atom is a generalized superposition of the excited state ($|e\rangle$) and the ground 
state ($|g\rangle$), $|\psi_{{a}}(0)\rangle = \cos{\frac{\theta}{2}}|e\rangle + \sin{\frac{\theta}{2}}|g\rangle$, where $0 
\leq \theta \leq \pi$.  $\theta=\pi/2$ corresponds to an initial coherent superposition of the two levels whereas 
$\theta=0$ corresponds to the initial state being the excited state, $|e\rangle$. The initial state of the two-level atom 
interacting with the 
coherent field can be written in the following way,
\begin{equation}
|\psi(0)\rangle = |\psi_{a}(0)\rangle \otimes |\psi_{f}(0)\rangle = \cos{\frac{\theta}{2}}|e, \alpha_N \rangle + 
\sin{\frac{\theta}{2}}|g,\alpha_N \rangle ,
\end{equation}
where $|\alpha_N \rangle = \prod_{j=1}^N |\alpha_j\rangle$. 

At any time $t$, the state vector of the total atom-field system evolves from the initial state according to (9). If we consider the initial atomic state to be $|e\rangle$ ($\theta=0$), the unitary time evolved state is
\begin{eqnarray}
|\psi(t)\rangle &=& U(t)|\psi(0)\rangle \nonumber \\
&=& \sum C_{n_j}^{\alpha}\{\cos{(\mathrm{\tilde{g}}t {\prod_{j=1}^N}\sqrt{n_j+1})}|e, \cal{N}\rangle \nonumber \\
&=& | e,\epsilon^+(t)\rangle + | g,\epsilon^-(t)\rangle ,
\label{eq4}
\end{eqnarray}
where $|\cal{N}\rangle = \prod_{j=1}^N |n_j\rangle $, $|\cal{N}+1\rangle = \prod_{j=1}^N |n_j+1\rangle$, and
\begin{eqnarray}
|\epsilon^+(t)\rangle &=& \sum C_{n_j}^{\alpha}\cos{(\mathrm{\tilde{g}}t {\prod_{j=1}^N}\sqrt{n_j+1})}|\cal{N}\rangle, 
\nonumber \\
|\epsilon^-(t)\rangle &=& -i \sum C_{n_j}^{\alpha} \sin{(\mathrm{\tilde{g}}t 
\prod_{j=1}^N\sqrt{n_j+1})}|\cal{N}+1\rangle.
\end{eqnarray}

\section{Reduced density operator for the atom and the field}
\label{mapping}

The time-evolved state of the two-level atom--multimode field interaction considered in section \ref{system} is a 
dynamical quantum state (\ref{eq4}) in a \textit{2}$\times\infty$ Hilbert space. However, the atom-field interaction is 
such that the dynamics of the composite pure system can be reduced onto an effective \textit{2}$\times$\textit{2} mapped state. This is 
due to the fact that the atomic transitions in the two-level system reduces the interacting field states to superpositions 
of the Fock states $|\cal{N}\rangle$ and $|\cal{N}+1\rangle$. We introduce two orthonormal basis states,
\begin{equation}
|\xi_1\rangle =(1/\delta_1)|\epsilon^+(t)\rangle ,
\end{equation}
and
\begin{equation}
|\xi_2\rangle = \frac{1}{\sqrt{1-|\emph{A}|^2}} \left[ 
\frac{|\epsilon^-(t)\rangle}{\delta_2}-\emph{A}\frac{|\epsilon^+(t)\rangle}{\delta_1}  \right] ,
\label{eq5}
\end{equation}
where $\delta_1=\sqrt{\langle\epsilon^+(t)|\epsilon^+(t)\rangle}$, 
$\delta_2=\sqrt{\langle\epsilon^-(t)|\epsilon^-(t)\rangle}$, and 
$\emph{A}=\langle\epsilon^+(t)|\epsilon^-(t)\rangle/\delta_1\delta_2$.\\\\
The time-evolved state in the orthonormal \textit{2}$\times$\textit{2} Hilbert space is given by
\begin{eqnarray}
 |\psi(t)\rangle = \delta_1 |e\rangle|\xi_1\rangle + \delta_2\sqrt{1-|\emph{A}|^2}|g\rangle|\xi_2\rangle + \delta_2 
 \emph{A} |g\rangle|\xi_1\rangle .
\label{eq6}
\end{eqnarray}
The density operator corresponding to the state (\ref{eq4}) using (\ref{eq6}) is
\begin{eqnarray}\nonumber
\rho(t)& =&  |\psi(t)\rangle\langle\psi(t)|\\
&=& \left(
        \begin{array}{cccc}
          \delta_1^2 & 0 & \delta_1 \delta_2 \emph{A} & \delta_1\delta_2\beta  \\
          0 & 0 &0 & 0 \\
          \delta_1 \delta_2 \emph{A}^* & 0 & \delta_2^2|\emph{A}|^2 & \delta_2^2\emph{A}\beta \\
         \delta_1\delta_2\beta & 0 & \delta_2^2\emph{A}^*\beta & \delta_2^2 \beta^2
        \end{array}
      \right),\\ \nonumber
\label{rho}
\end{eqnarray}
where $\beta= \sqrt{1-|\emph{A}|^2}$, $\delta_i=\delta_i^*$ (i = 1, 2), and $\emph{A}=-\emph{A}^*$. The density matrix can 
be numerically calculated by evaluating $\delta_i$ (i=1,2) and $\emph{A}$, provided the properties of the interacting 
coherent field are known. The \textit{4}$\times$\textit{4} mapped density matrix has rank 3. Similar rank-3 density 
matrices are also seen in other forms of atom-photon interaction where there is a mapping of states from the atomic to the 
photonic subsystem \cite{aop}.
The atomic and the field subsystems at time $t$ can be obtained as reduced density matrices by tracing the correlated 
density matrix, $\rho(t)$ over the field and the atomic variables, respectively.

The mapping of the (\textit{2}$\times \infty$) infinite level system onto an effective \textit{2}$\times$\textit{2} system 
is possible for closed pure state dynamics of the atom-photon system, where the effective Schmidt rank of the composite 
system is governed by the lower dimension of the two subsystems. 
For dissipating open systems, the mixed state dynamics can be reduced by optical truncation of the infinite dimensional field using quantum engineered operations called quantum scissors \cite{sciss}.
Let us consider a dissipative nonunitary evolution of the density matrix under phase damping of the atomic system. The 
environment-induced decohering open system can be written in its Lindblad form \cite{car} as
\begin{equation}
\dot{\rho}(t)=-\frac{i}{\hbar}[H, \rho(t)] + \frac{\gamma}{2}\left(\sigma_z \rho(t) \sigma_z - \rho(t)\right),
\label{lind}
\end{equation}
where $\gamma$ is the phase decoherence parameter, and $\sigma_z = \left|e\rangle\langle e\right| - \left|g\rangle\langle 
g\right|$. The nonunitary, mixed state evolution represented by equation (\ref{lind}) does not allow solutions that can be mapped from 
the infinite level atom-photon interaction to an equivalent $\textit{2}\times\textit{2}$ state. 
The infinite-level problem for mixed states can be reduced using suitable optical truncation of the interacting field. 
Using quantum scissors, the $N$-mode input coherent state can be written as
\begin{equation}
|\psi_f'(0)\rangle = \frac{1}{\mathcal{N}}\prod_{j=1}^{N} |\alpha'_j\rangle = 
\frac{1}{\mathcal{N}}\sum_{n_1,n_2,...,n_N=0}^{k} \mathcal{C}_{n_j}^\alpha \prod_{j=1}^N |n_j\rangle ,
\end{equation}
where the infinite summation over all number states has been truncated to summation over $k$ states. $\mathcal{N}$ is the 
new normalization constant. This reduces the infinite dimensional atom-photon interaction to a reduced $\textit{2}\times\textit{k}$ composite state problem.
For non-dissipative pure state dynamics, it can be shown that, for a suitable choice of $k$, the dynamics of the truncated $\textit{2}\times\textit{k}$ state is equivalent to that of the mapped $\textit{2}\times\textit{2}$ system, derived in (\ref{eq6}). 
For a single-mode field, truncated to $k$ states, the following simplifications can be made:
\begin{eqnarray}
[H_{int},\rho(t)] &=& \sum_{i=0}^{k-1} \sqrt{i+1}~ \left(\mathrm{\tilde{g}}~ |e,i\rangle\langle g, i+1| + \mbox{H.c.} 
\right), \\
\frac{\gamma}{2}\left(\sigma_z \rho(t) \sigma_z - \rho(t)\right) &=& \sum_{m=k/2+1}^k ~\sum_{l=0}^{k/2} - ~\gamma 
\left(|e,l\rangle\langle g,k| + \mbox{H.c.} \right) .
\end{eqnarray}
In the composite atom--number state basis, the density matrix $\rho(t)$ can be written as
\begin{eqnarray}
\rho(t) &=&  \sum_{i=0}^{k/2}~\sum_{j=0}^{k/2} c_{ij}(t) |e,i\rangle\langle e,j| + \sum_{i=k/2+1}^{k}~\sum_{j=k/2+1}^{k} 
c_{ij}(t) |g,i\rangle\langle g,j| \nonumber\\
&& + \sum_{i=0}^{k/2}~ \sum_{j=k/2+1}^{k} c_{ij}(t) |e,i\rangle\langle g,j| + \mbox{H.c.}
\label{rho2}
\end{eqnarray}
Using the form of the density matrix (\ref{rho2}) in the atom--number state representation, the Lindblad equation 
(\ref{lind}) can be written in the form
\begin{equation}
\dot{\rho}(t) = \mathcal{L}\rho(t),
\label{fin}
\end{equation}
where $\mathcal{L}$ is the superoperator of the dynamic nonunitary mapping. Solving equation (\ref{fin}) for a known 
initial state $\rho(0)$, and a suitable truncation number $k$, we can obtain the final dephased state of the atom-photon 
interaction (with dissipation). This final state, $\rho(t)$, can then be used to calculate the quantum correlation properties of the 
composite system.

\section{Correlation dynamics of the atom-field system}
\label{results}

The quantum correlation properties of the reduced \textit{2}$\times$\textit{2} atom-field system (\ref{eq6}), obtained by mapping the infinite-dimensional pure state interaction, can be seen by studying the dynamics of the entanglement of formation. All entropic measures such as quantum discord reduce to EOF for pure quantum states. For consistency in comparing with mixed state dynamics, we also consider the evolution of the logarithmic negativity, which is an upper bound for the EOF for pure states. Using single- and two-mode fields, we check and compare the dynamics of EOF [$E(\rho)$] and LN [$L(\rho)$], defined in section \ref{correlation}.

\begin{figure}[ht]
\label{fig:1}
\begin{center}
\epsfig{figure=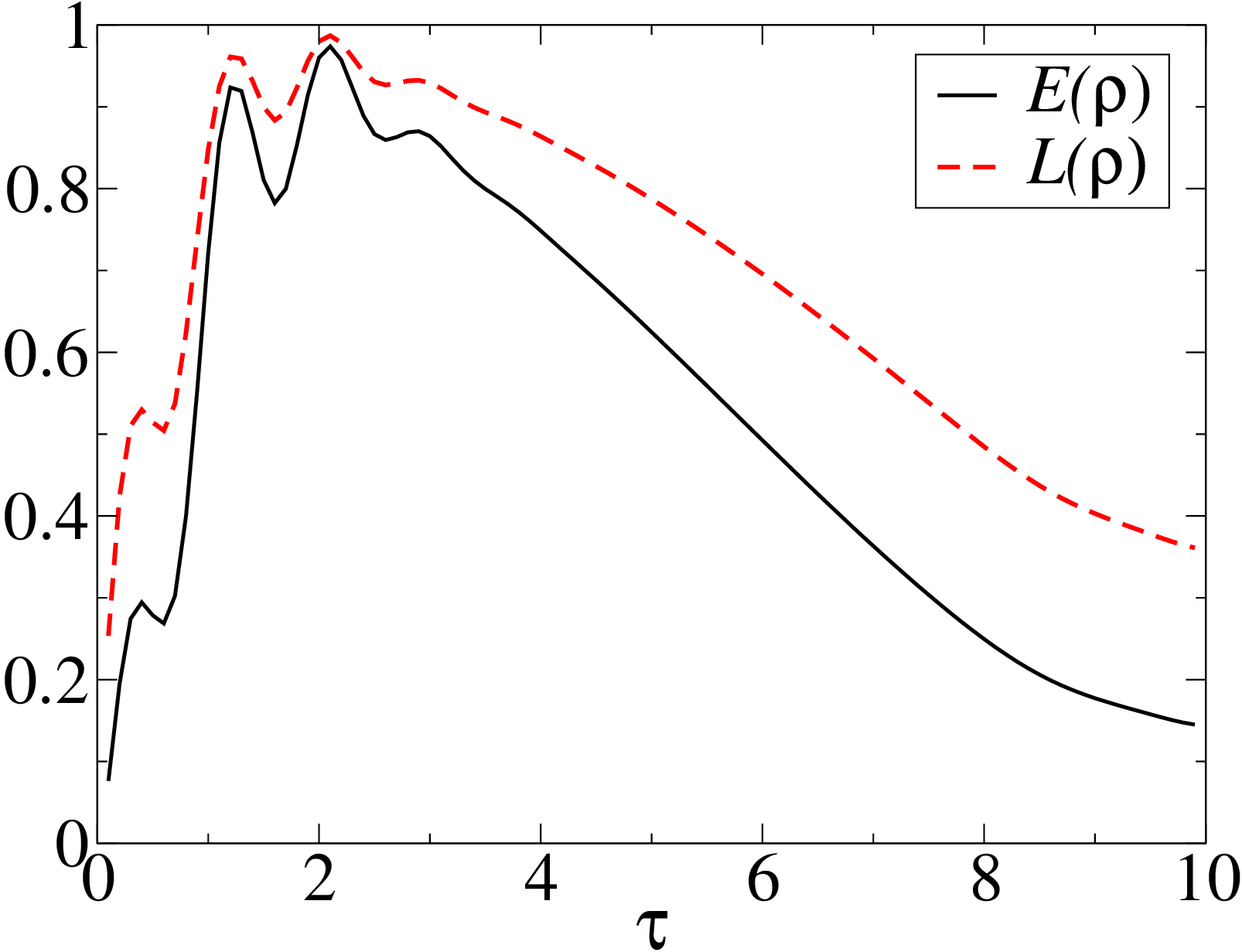, height=.18\textheight, width=.36\textwidth, angle=-0}
\caption{Time evolution of quantum correlations, EOF [$E(\rho)$] and LN [$L(\rho)$], in the two-level atom interacting with a single-mode field, with average photon number, $\bar{n}=10$. $\tau \equiv \tilde{\mathrm{g}}t$ is the scaled dimensionless time. The dynamics of EOF (black continuous line) and LN (red dashed line) show that the system is highly correlated and is maximally entangled around $\tau \approx$ 2.0. LN forms a tight upper bound on EOF, which for pure states is equal to the entropy of entanglement. Measures such as quantum discord are equal to EOF for the pure state dynamics. }
\end{center}
\end{figure}
%

Figure 1 shows the evolution of the EOF and LN as a function of the scaled evolution time ($\tau \equiv 
\tilde{\mathrm{g}} t$), for a single-mode initial coherent field, with average photon number, $\bar{n}$ = 10. We observe 
that reduced atom-photon composite system is highly correlated. At times, $\tau \approx$ 2.0, the system is maximally 
entangled. The bipartite correlation properties of the infinite-dimensional atom-photon interaction is captured by the 
reduced dynamics of the mapped system. Since the mapping retains the unitary evolution or pure state dynamics of the 
composite system, all entropic correlations of the bipartite system can be captured using the EOF. 

\begin{figure}[ht]
\label{fig:2}
\begin{center}
\epsfig{figure=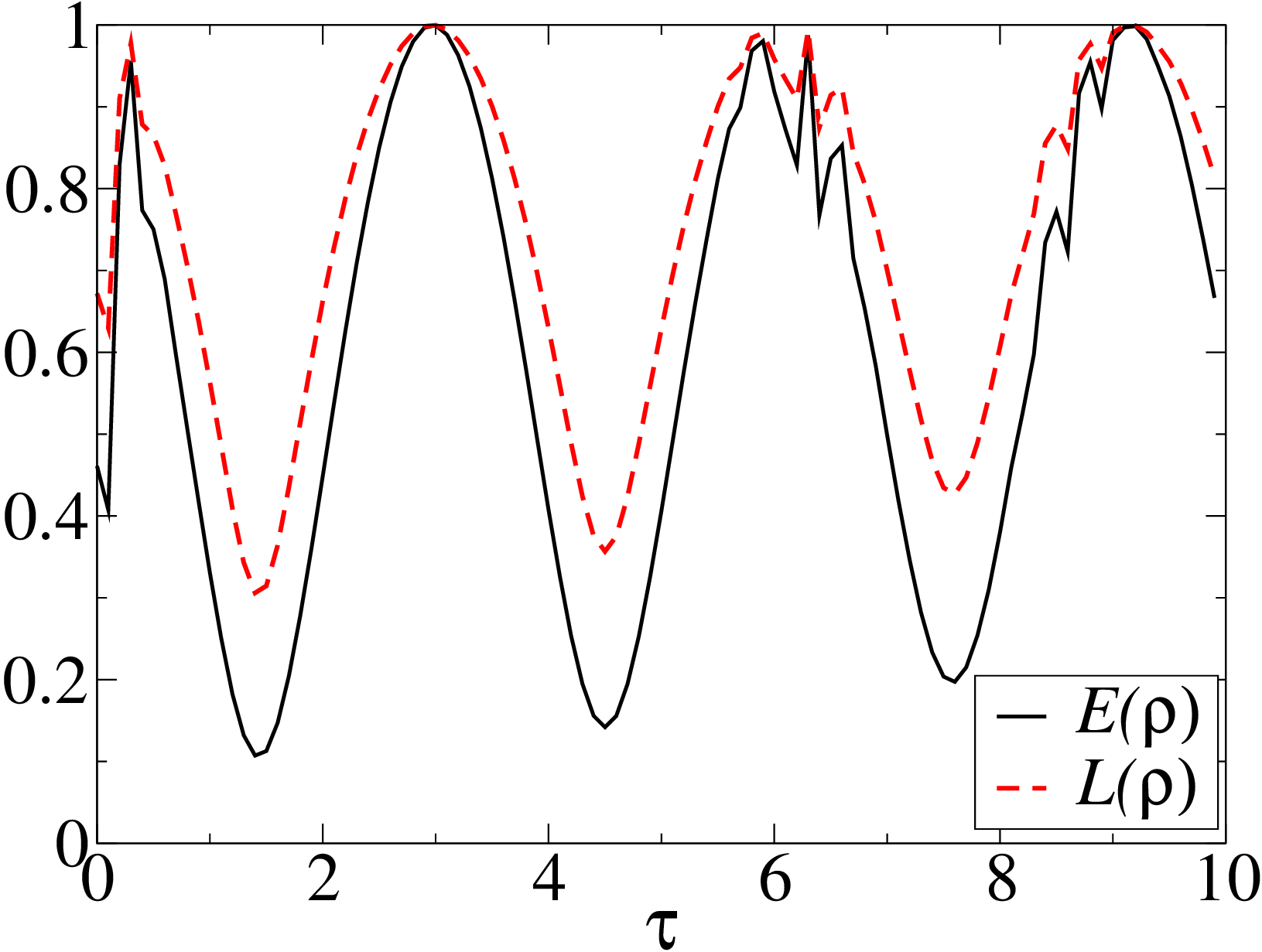, height=.18\textheight, width=.36\textwidth, angle=-0}
\caption{Time evolution of EOF [$E(\rho$)] and LN [$L(\rho$)] in the two-level atom interacting with a two-mode field, with average photon numbers, $\bar{n}_1=10$, and $\bar{n}_2=10$. LN (red dashed line) forms an upper bound on EOF (black continuous line), which is equal 
at all times to QD.}
\end{center}
\end{figure}

In figure 2, we plot the time evolutions of the EOF and LN, for a two-mode initial field with average photon numbers, 
$\bar{n}_1 = \bar{n}_2$ = 10, as a function of the scaled evolution time ($\tau$).
%
The EOF, as expected, is bounded above by LN, similar to the correlation bound observed for the case of initial single-mode field.
The oscillatory and discontinuous nature of the quantum correlations in the two-mode case is due to the fact that the collapse and revival of Rabi oscillations occur rapidly in short intervals \cite{gou}. QD is equal to EOF at all times since the dynamics is 
unitary (pure). Hence, the increase in the number of modes does not qualitatively affect the dynamics of entanglement in 
the atom-photon system.


We next consider the pure state dynamics of the optically truncated, quantum scissored state given by relation 
(\ref{rho2}), with $\gamma$ = 0. We probe different values of $k$ for which the pure, optically truncated \textit{2}$\times$\textit{k} states attempt to simulate the dynamics of the mapped \textit{2}$\times$\textit{2} system given by relation (\ref{eq6}). 
Figure 3 shows the evolution of the EOF for the infinite optical field ($k = \infty$) mapped to the reduced \textit{2}$\times$\textit{2} state (\ref{eq6}) in comparison to the optically truncated \textit{2}$\times$\textit{k} systems (\ref{rho2}), corresponding to $k$ = 12, 16, and 20, for a single-mode field with average photon number, $\bar{n}$ = 10. We observe that at $k$ = 20, the optically truncated, quantum scissored state effectively simulates the correlation dynamics of the reduced infinite system. For calculations 
involving dissipative mixed state dynamics in the following section, we have used $k$ = 30.

\begin{figure}[ht]
\label{fig:3}
\begin{center}
\epsfig{figure=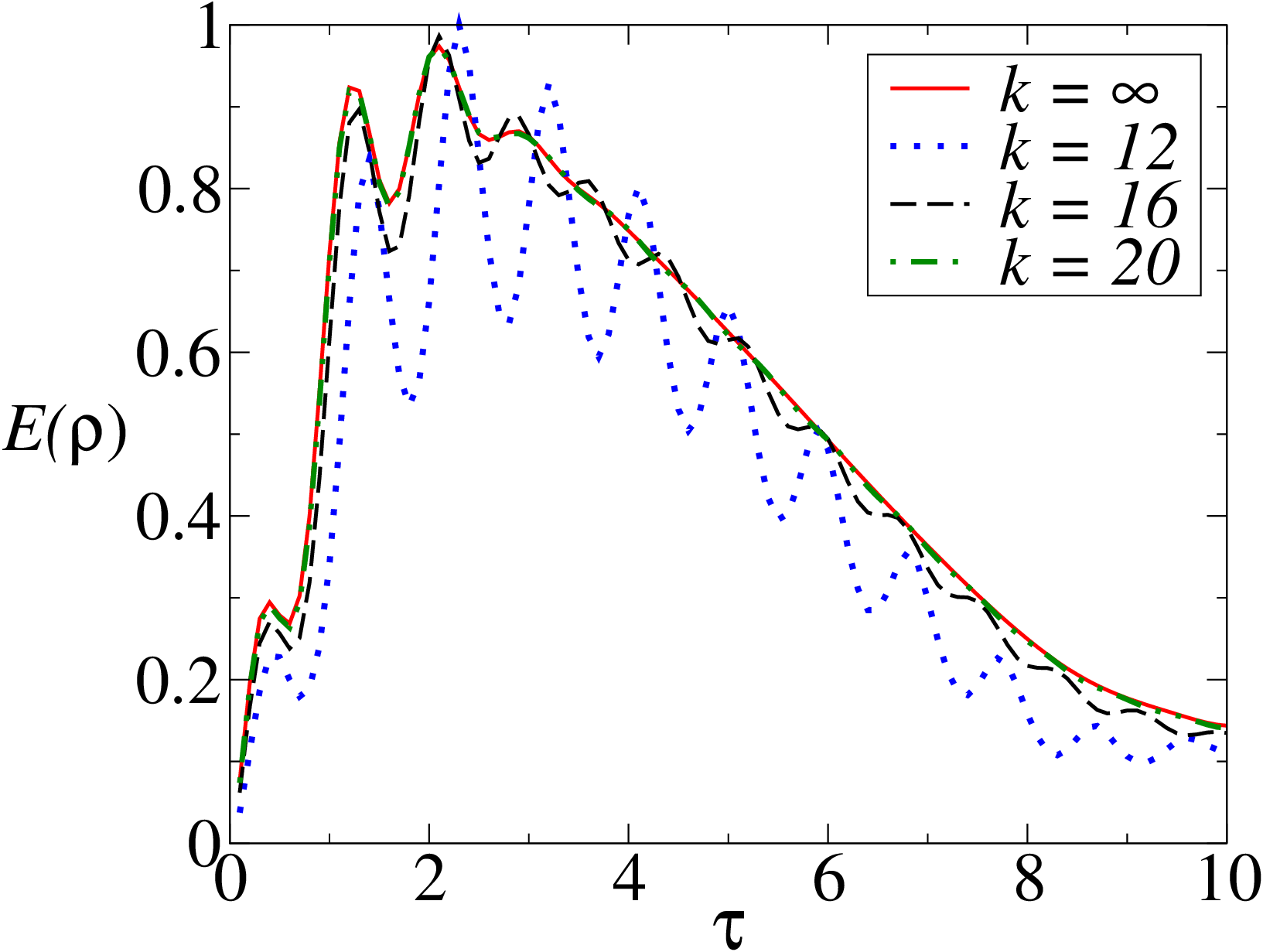, height=.2\textheight, width=.4\textwidth, angle=-0}
\caption{Dynamics of EOF in the two-level atom interacting with a single-mode field, with average photon number 
$\bar{n}=10$, for different values of optical truncation number $k$. The nontruncated mapped system corresponds to $k=\infty$ (red 
continuous line). The truncated states are shown by blue dotted ($k = 12$), black dashed ($k = 16$), and green dot-dashed ($k = 
20$) lines. We observe that the evolution of EOF for $k = 20$ truncated state overlaps with that for the nontruncated mapped system.
}
\end{center}
\end{figure}

\section{Quantum correlation dynamics under atomic phase damping}
\label{damp}

The atomic phase damping in the atom-photon interaction can be suitably studied by the dissipative nonunitary 
transformation of the composite density matrix governed by the Lindblad equation (\ref{lind}). Such phase 
damping of the atomic subsystem may arise due to elastic collisions in atomic vapor in trapped systems \cite{AKR3}. The 
evolution of the damped system under nonunitary transformation is mixed, and the infinite-dimensional atom-photon interaction is reduced to a \textit{2}$\times$\textit{k} state using optical truncation (\ref{rho2}). For \textit{2}$\times$\textit{k}-dimensional mixed states, logarithmic negativity is a computable measure of distillable entanglement. In this regime, information-theoretic measures such as quantum discord are no longer equal to bipartite entanglement. In this section, we study the effect of atomic phase damping on the evolution of LN and QD, in the optically truncated, finite-dimensional states. The robustness and sudden death of distillable entanglement in arbitrary dimensional mixed states have been partly investigated in \cite{dsd}.

\begin{figure}[ht]
\begin{center}
\subfigure[]{\includegraphics[width=5.cm]{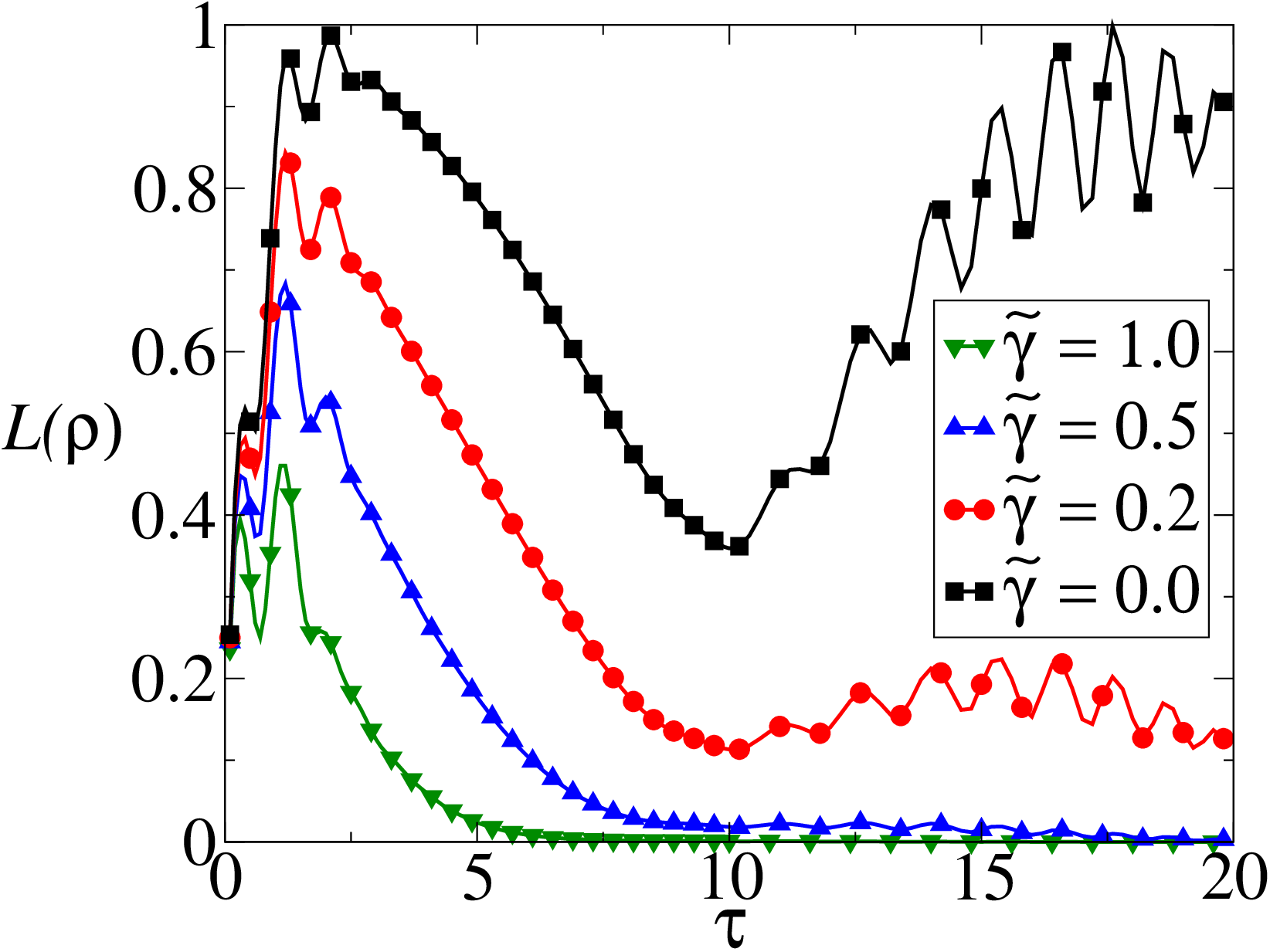}}\hspace{0.3cm}
\subfigure[]{\includegraphics[width=5.cm]{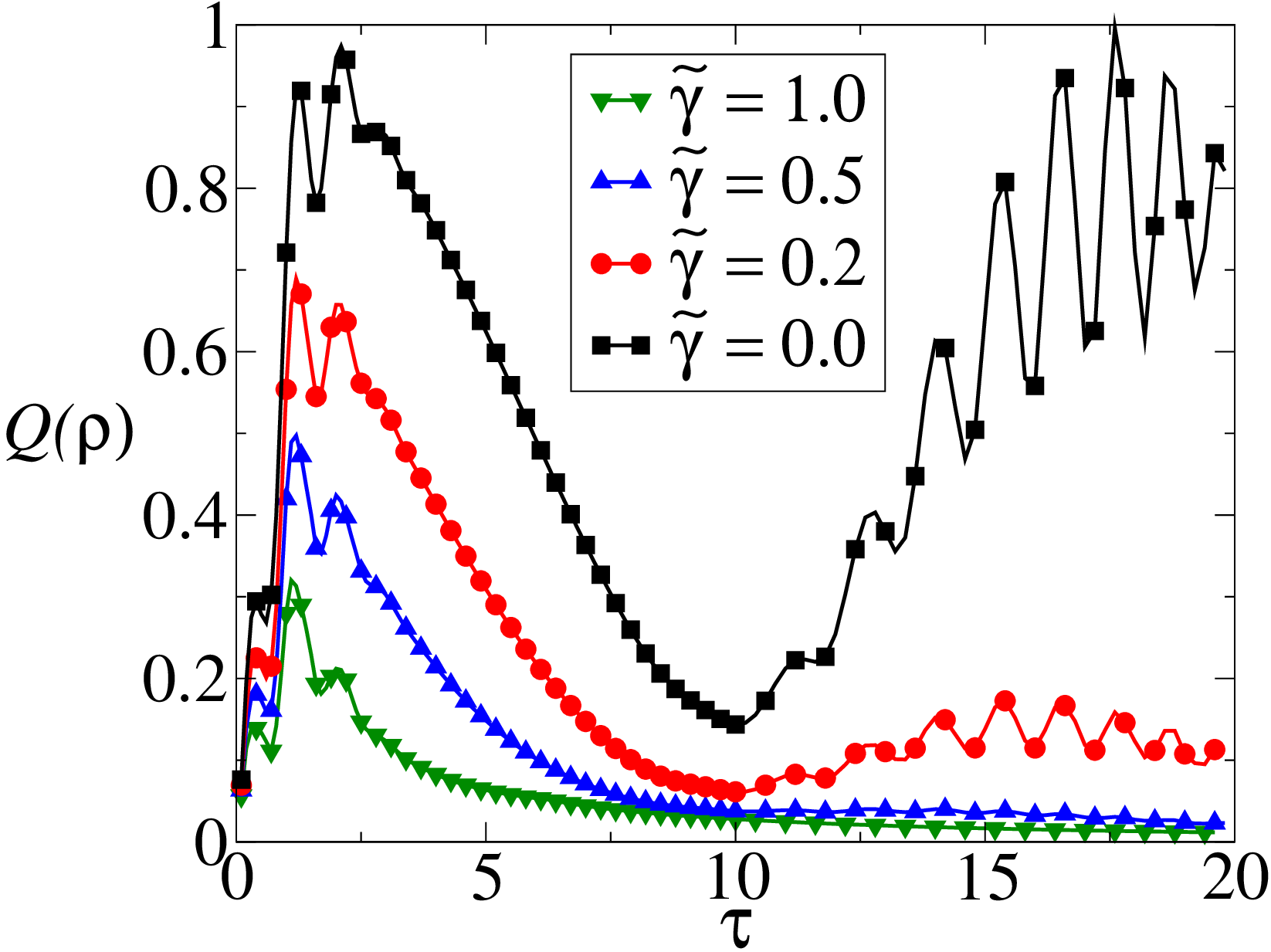}}
\caption{Dynamics of (a) LN [$L(\rho)$] and (b) QD [$Q(\rho)$],
for a two-level atom interacting with a single-mode field, with average photon number $\bar{n}=10$, for different values of the atomic phase damping: $\tilde{\gamma}$ = 1.0 (green down triangles), $\tilde{\gamma}$ = 0.5 (blue up triangles), $\tilde{\gamma}$ = 0.2 (red circles), and $\tilde{\gamma}$ = 0.0 (black squares). $\tilde{\gamma}$ = $\gamma / \tilde{\mathrm{g}}$, and $\tau$ = $\tilde{\mathrm{g}}$t are the scaled, dimensionless dephasing parameter and time,
respectively. We observe that with increase in the scaled dephasing parameter $\tilde{\gamma}$, the quantum correlations 
decay faster with increase in the scaled time ($\tau$).}
\end{center}
\label{fig:4}
\end{figure}

Figure 4 shows the dynamics of the quantum correlation measures, LN and QD, under the effect of atomic dephasing. QD is numerically evaluated using relation (\ref{QD}) with projective measurement done on the atomic subsystem. We observe that the quantum correlations steadily decay with increase in the scaled dephasing parameter $\tilde{\gamma}$ (= $\gamma / \tilde{\mathrm{g}}$) as it evolves in time $\tau$. Further, we observe that as time is increased, the LN in the system decreases to values close to zero (around $\tau \approx$ 10) for values of dephasing parameter $\tilde{\gamma}$ = 0.5 and 1.0. This implies that the distillable entanglement in the system vanishes at times ($\tau \ge$ 10) for high values of $\tilde{\gamma}$. However, the information-theoretic measure, QD, does not vanish with increasing time, converging at a low value. Further, for zero or low damping, there is distinct oscillatory characteristic for all the measures. This property is reminiscent of the Rabi oscillations in the system.

To study the robustness of the correlation measures, we study the decay of quantum correlations for highly correlated 
initial states. From figures 1 and 3, we see that the composite atom-photon system is close to maximal entanglement at 
$\tau \approx$ 2.0 for undamped, pure unitary evolutions ($\tilde{\gamma}$ = 0.0). We apply a protocol whereby the decay term leading to nonunitary evolution is activated at $\tau$ = 2.0. Hence, the system evolves unitarily to a near maximally 
entangled state till $\tau$ = 2.0, when the atomic phase damping is applied ,and the system nonunitarily decays to lower 
correlated states. The decay of the different measures reflects the robustness of the quantum correlations under atomic 
phase damping.

\begin{figure}[ht]
\begin{center}
\subfigure[]{\includegraphics[width=5.cm]{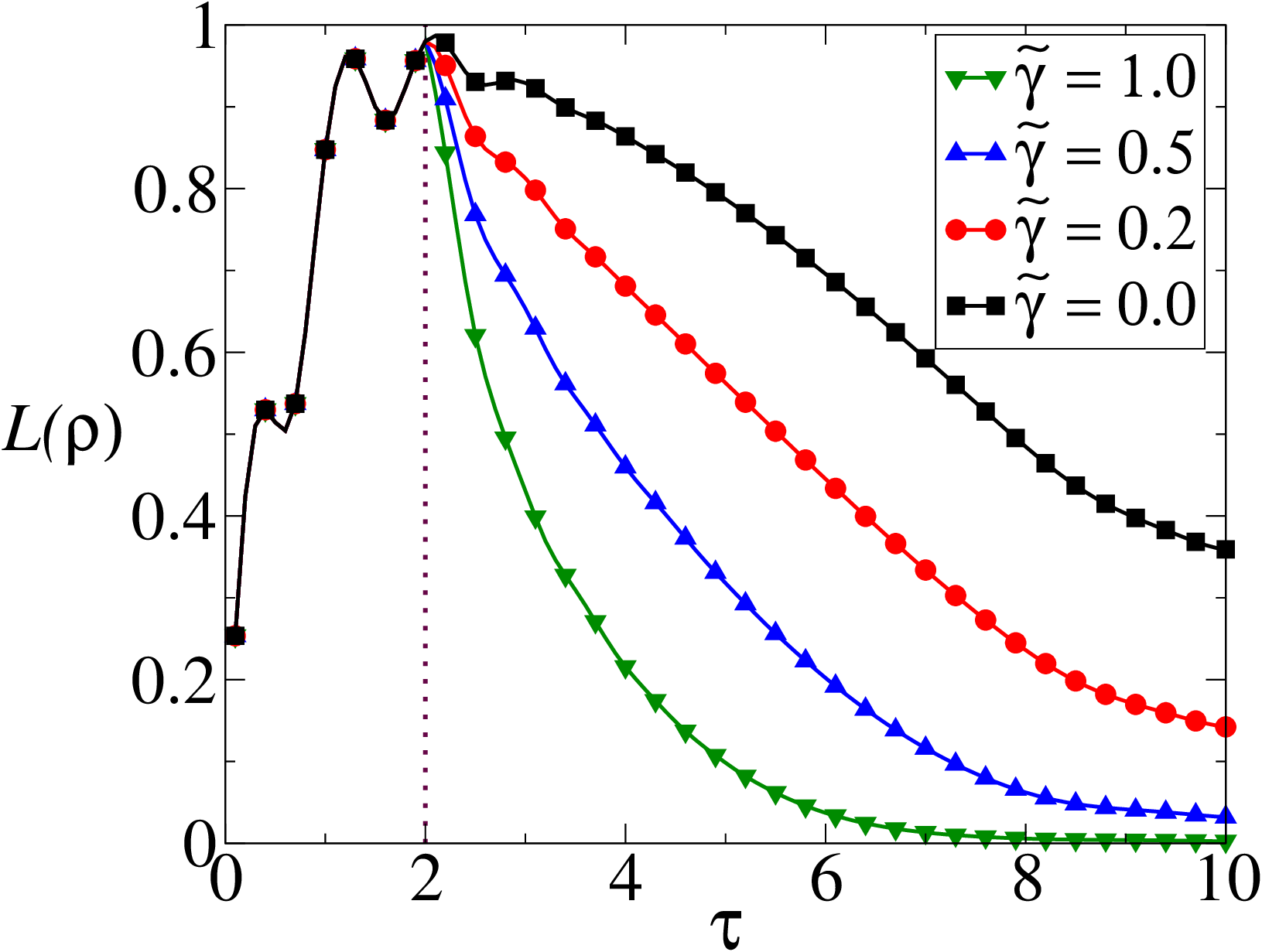}}\hspace{0.3cm}
\subfigure[]{\includegraphics[width=5.cm]{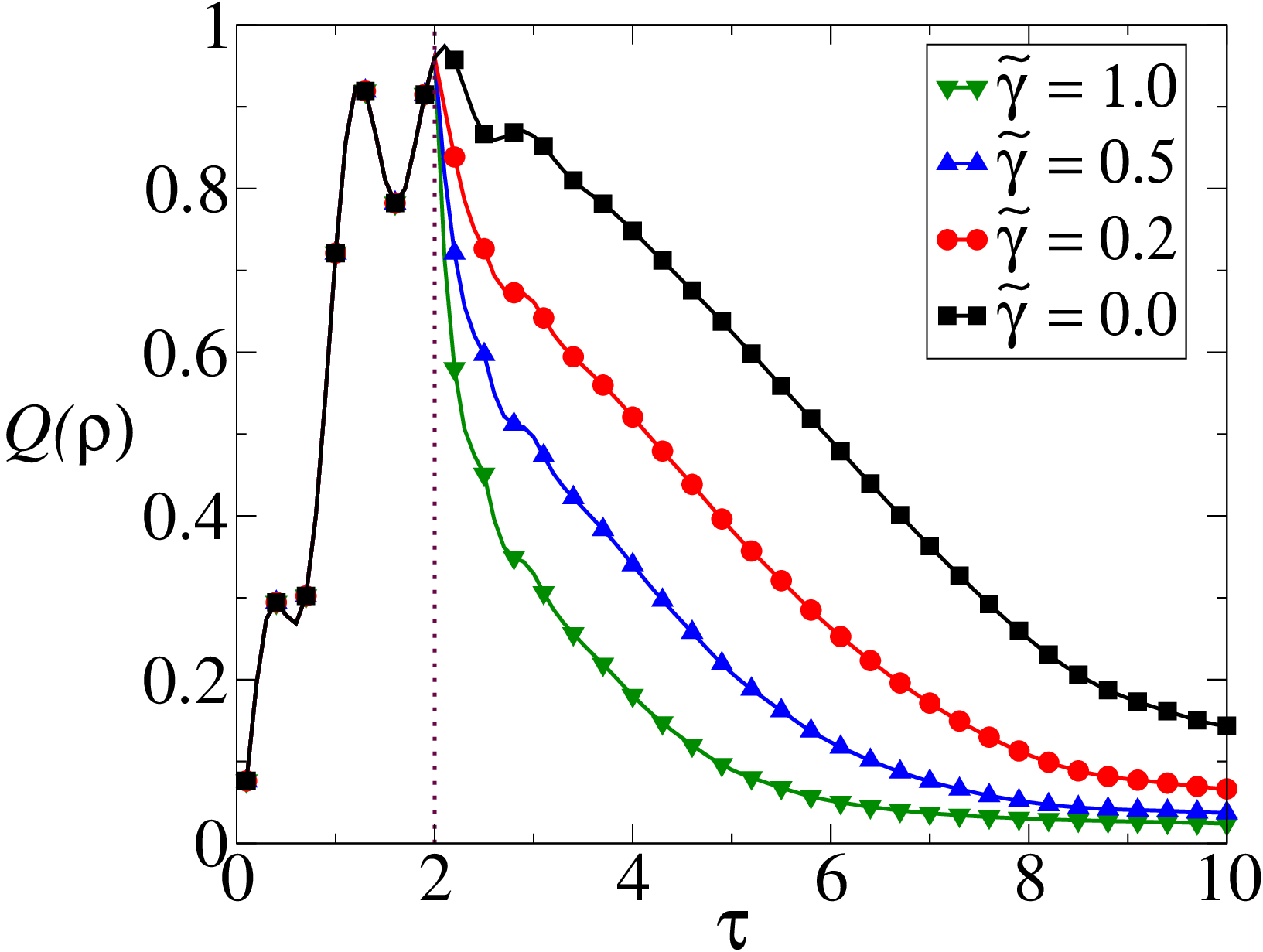}}
\caption{The same as figure 4, but now with atomic dephasing applied at time, $\tau$ = 2.0 (maroon dotted vertical line) onwards.}
\end{center}
\label{fig:5}
\end{figure}

From figure 5, we observe that even for highly entangled initial states, the decay of quantum correlation possesses 
certain characteristic features. For high values of scaled atomic dephasing ($\tilde{\gamma}$ = 1.0), the LN reduces to 
values close to zero. Hence, the system possesses no distillable entanglement. However, the decay of QD ensures 
that some residual amount of quantum correlations always remains between the atomic and photonic subsystems.
\begin{figure}[b]
\label{fig:6}
\begin{center}
\epsfig{figure=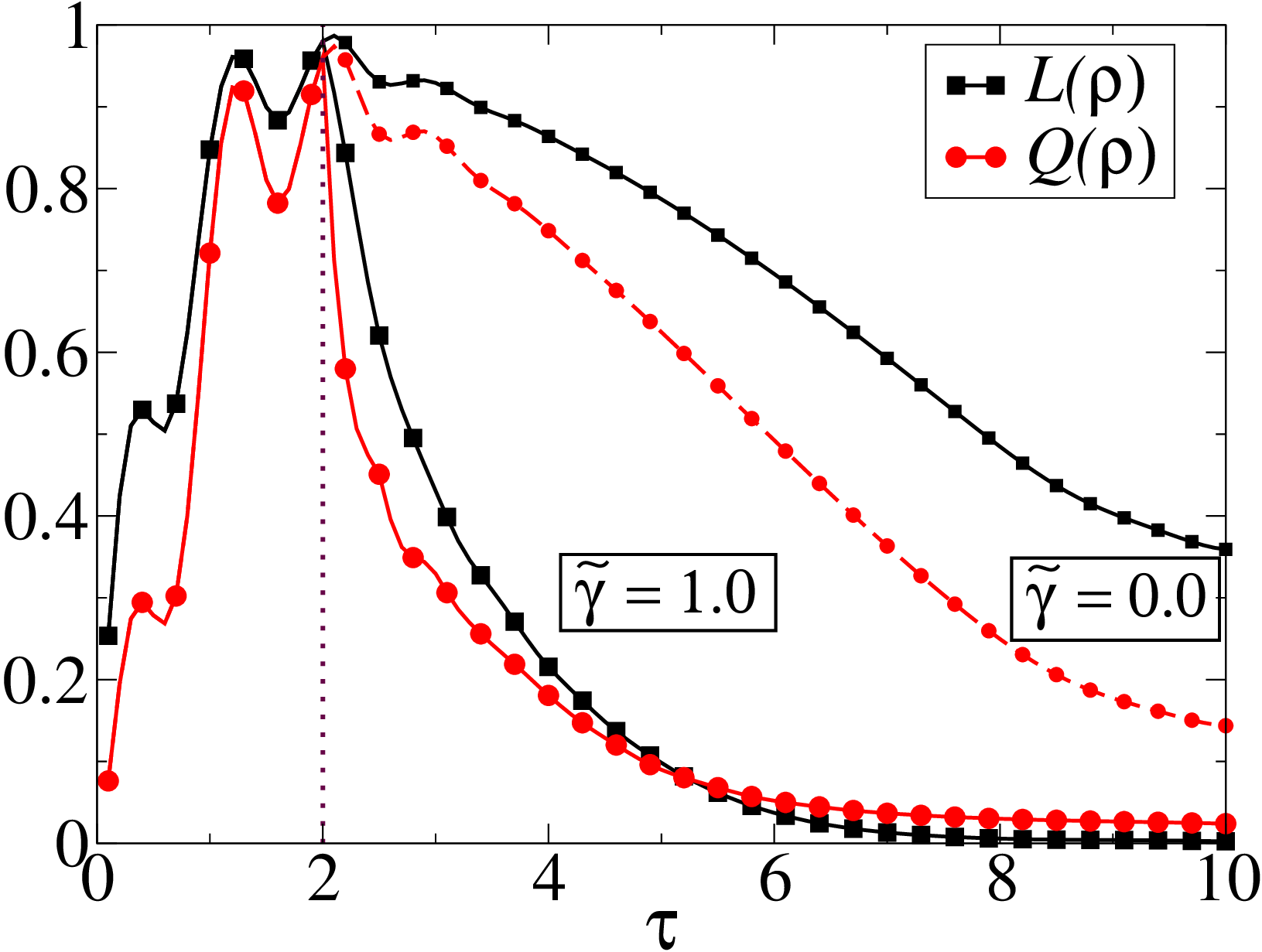, height=.18\textheight, width=.36\textwidth, angle=-0}
\caption{Comparison of LN (black squares) and QD (red circles)
for the undamped ($\tilde{\gamma}$ = 0.0) and the damped ($\tilde{\gamma}$ = 1.0) two-level atom interacting with a single-mode field, with average photon number $\bar{n}$ = 10. The atomic dephasing is applied for times, $\tau >$ 2.0 (maroon dotted vertical line). We observe that LN decays more rapidly and is less robust compared to QD for increasing $\tau$ in the presence of atomic dephasing.}
\end{center}
\end{figure}

Figure 6 shows a neat comparison between the different quantum correlation measures for an undamped ($\tilde{\gamma}$ = 
0.0) and a damped ($\tilde{\gamma}$ = 1.0) two-level atom interacting with a single-mode coherent input with average 
photon number, $\bar{n}$ = 10. The atomic dephasing is applied at time, $\tau$ = 2.0. Hence, the effective initial field 
at the onset of damping is almost maximally entangled. For the undamped system, the evolution is unitary and the system is 
always pure during the dynamics. Hence, QD is equal to the entanglement of formation and LN is an upper bound on 
these entropic measures. In case of the damped system, the system evolution is nonunitary, dissipative and mixed due to 
atomic dephasing induced by the environment. Hence, there is imminent decay of the quantum correlations. Since QD 
is no longer entropically equivalent to entanglement, the tight bound of LN over QD vanishes.
Figure 6 shows that QD is more robust to dephasing than LN. We observe that LN decreases at a greater rate and is 
quenched with increase in time. QD decreases at a relatively slower late and is never completely quenched in the 
dynamics. This shows that QD is robust even at dissipation rates that lead to sudden death of distillable entanglement.

\begin{figure}[ht]
\label{fig:7}
\begin{center}
\epsfig{figure=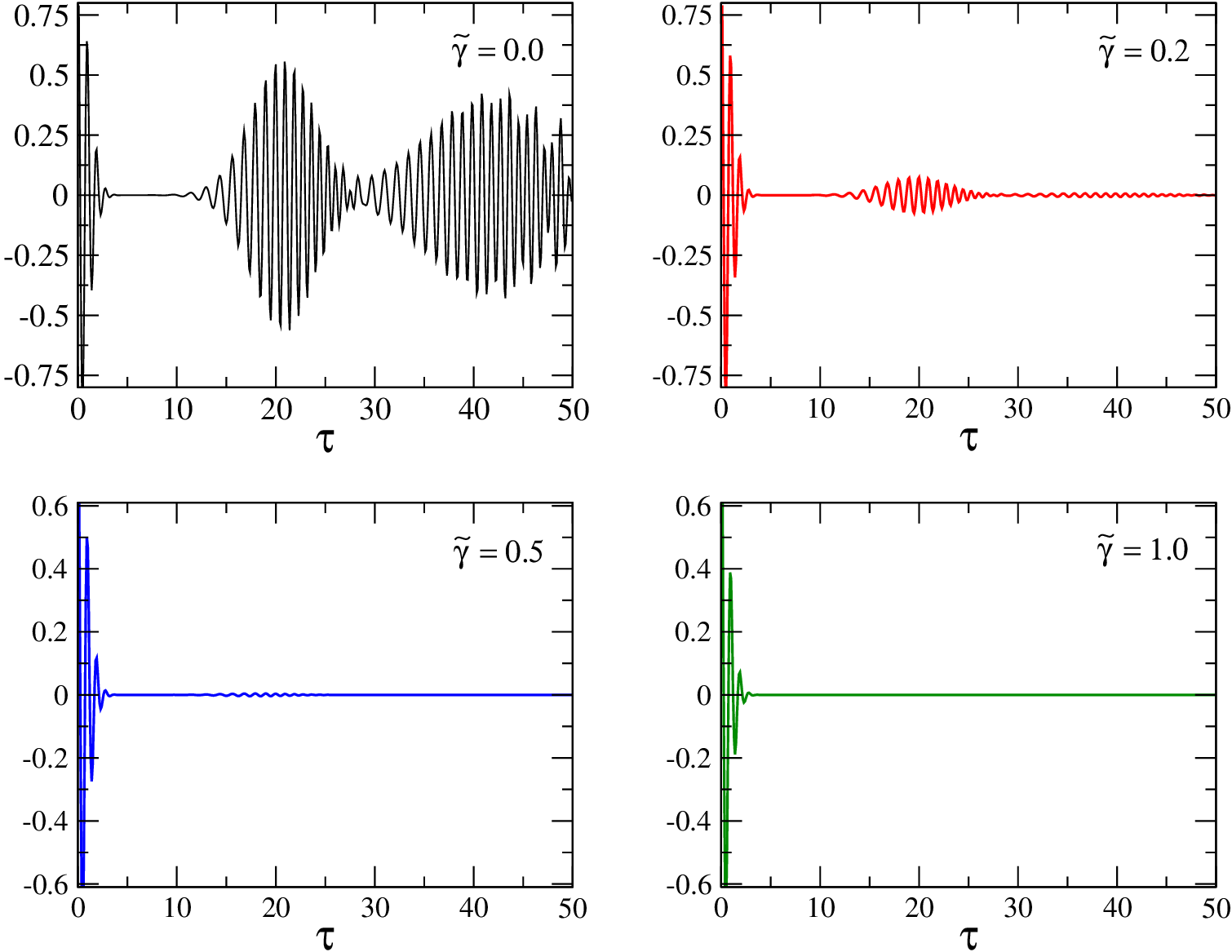, height=.4\textheight, width=0.8\textwidth, angle=-0}
\caption{The collapse and revival of Rabi oscillations in a two-level atom interacting with a single-mode field, with 
average photon number $\bar{n}$ = 10, for different values of atomic dephasing ($\tilde{\gamma}$). We observe that for 
greater damping, there is no revival, which is reminiscent of the behavior of the quantum correlations.}
\end{center}
\end{figure}

In figure 7, we show the effect of atomic phase damping on the collapse and revival of Rabi oscillations in the system. 
It is known that the undamped JC model exhibits distinct revival after the collapse of the initial system, which is the 
hallmark of the nonclassical nature of photons. We observe that the dynamics of the quantum correlations in the damped 
atom-photon system complement the collapse and revival of the Rabi oscillations for a single-mode initial coherent field 
with photon number, $\bar{n}$ = 10. From figure 4, it is evident that the oscillation in the correlation of the undamped 
($\tilde{\gamma}$ = 0.0) system, with increasing scaled time ($\tau$), fades for higher dephasing parameter, 
$\tilde{\gamma} >$ 0.0. This is evident in the gradual decay of the revival of Rabi oscillations with increasing 
$\tilde{\gamma}$. For $\tilde{\gamma}$ = 1.0, the revival of Rabi oscillations is almost absent, and this is manifested in 
the close to zero values of LN and QD for increasing time $\tau$. This is an interesting qualitative relation as the 
Rabi oscillations correspond to the difference in atomic level populations whereas the quantum correlations correspond to 
the coherence elements of the density matrix. The nonclassical feature of the interaction contained in the density matrix 
elements is highlighted in the qualitatively similar oscillations of the quantum correlations and population inversion. 
Hence, the nonclassical revival of Rabi oscillations is ingrained in the atom-field quantum correlations.

\section{Discussions}
\label{dissc}

A generalized characterization and study of atom-field interaction is of great importance for the practical implementation 
of the field of quantum information and computation. A simple generalized JC model interacting with multimode coherent 
state input is able to generate highly  correlated quantum systems that may be put to use for performing basic quantum 
protocols. The important qualification for such a model is the fact that an interaction between a discrete two-level 
atomic system and a continuous variable multimode quantum field can be mapped into an effective finite-dimensional bipartite system 
interaction that contains the dynamics of the highly correlated atom-field system. Such systems can also be suitably engineered using optical truncation to simulate low-dimensional dynamics. This enables one to study dissipation induced by atomic dephasing in the atom-photon interaction. 
We observe that during the course of the temporal evolution, the atom-field correlated system is maximally entangled at certain scaled times. For dissipative mixed state dynamics, we observe that certain information-theoretic quantum correlation measures are more robust than mixed state entanglement. The oscillations of the atom-field correlations are closely related to the observed phenomena of the collapse and revival of Rabi oscillations in such systems.

An important aspect of studying these models is the distinct possibility of generating such bipartite interactions in the 
laboratory for investigating and applying these correlated systems in future quantum tasks. Harnessing the power of quantum correlations with respect to quantum tasks and protocols of the future depends strongly on the ability to control, manipulate and address quantum systems with high precision and coherence. Understanding the basic dynamics of quantum correlations in simple atom-field interactions is of utmost importance in the realization of such controlled quantum systems.


\ack
HSD thanks University Grants Commission, India, and AC thanks National Board of Higher Mathematics, Department of Atomic 
Energy, Government of India, for financial support.

\end{document}